\documentclass[reprint, amsmath, amssymb, aps, superscriptaddress]{revtex4-1}
\usepackage{graphicx}% Include figure files
\usepackage{dcolumn}% Align table columns on decimal point
\usepackage{bm}% bold math
\usepackage{hyperref}% add hypertext capabilities
\usepackage{float}
\usepackage{color}

\newcommand{\rom}[1]{\uppercase\expandafter{\romannumeral #1\relax}}

\begin{document}

\title{Complex Phase Diagram of Doped XXZ Ladder: Localization and Pairing}
  \author{Rong-Yang Sun}
  \affiliation{Institute for Advanced Study, Tsinghua University,
  Beijing, 100084, China}
  \author{Zheng Zhu}
  \affiliation{Department of Physics, Harvard University, Cambridge, MA, 02138,
  USA}
  \author{Zheng-Yu Weng}
  \affiliation{Institute for Advanced Study, Tsinghua University,
  Beijing, 100084, China}
  \date{\today}

\begin{abstract}
  How the ground state nature can be dramatically changed by the distinct underlying spin correlation is a central issue of doped Mott insulators. The two-leg XXZ ladder provides a prototypical spin background, which can be tuned from a long-range N\'{e}el order to a short-range ``spin liquid'' via the superexchange anisotropy, giving rise to a complex phase diagram at finite doping. By density matrix renormalization group method, we show that although the charge is always self-localized in the N\'{e}el ordered phase, a second insulating phase emerges, in which the doped holes become paired but remain localized while the transverse spin-spin correlation reduces to short-ranged one to make the N\'{e}el order classical. Only when the N\'{e}el order totally disappears by further reducing anisotropy, does the pairing become truly coherent as characterized by a Luther-Emery state. In sharp contrast, the pairing is totally absent in the in-plane ferromagnetic XXZ regime, where a direct transition from the charge self-localization in the N\'{e}el ordered phase to a Fermi-gas-like state in the spin liquid phase is found. A consistent physical picture is briefly discussed.
\end{abstract}

\maketitle

\emph{\label{sec:intro}Introduction.---}As one of the simplest models describing the doped Mott insulator, the $t$-$J$ model has attracted intense attention due to its potential to characterize systematically a complex phase diagram composed of a long-range N\'{e}el state, pseudogap and superconducting (SC) states, etc., as a function of doping \cite{lee2006doping}.  Such a model can be accurately investigated by density matrix renormalization group (DMRG) method \cite{white1992density} in quasi one-dimensional (1D) square lattice cases (i.e., the ``ladders''), which does exhibit quasi-1D SC behavior \cite{poilblanc1995spin,siller2002transition,jiang2018superconductivity,jiang2019critical,jiang2019ground} known as the Luther-Emery (LE) state, characterized by an exponential decay of spin correlations and power-law decays of SC and charge density wave (CDW) correlations \cite{luther1974backward,jiang2019critical,seidel2005luther}. As a matter of fact, an \emph{intrinsic} pairing force has been clearly identified even for two doped holes in the two- and four-leg $t$-$J$ ladders\cite{zhu2014nature,sun2019localization}. Here the half-filling isotropic spin background in the ladder is already a ``spin liquid'' without long-range N\'{e}el order \cite{dagotto1996surprises}, which would seem to support the resonance valence bond (RVB) idea of Anderson for the SC origin out of a pure repulsive interaction\cite{anderson1987resonating,baskaran1987resonating}.

However, given the same ``RVB'' or gapped spin background, the LE ground state can be replaced by a Luttinger-liquid-like state with the exponent very close to that of the free Fermi gas (FG) \cite{jiang2019critical} at finite doping, \emph {if} the hopping term is modified to result in the so-called  $\sigma\cdot$$t$-$J$ ladders \cite{zhu2013strong,zhu2015quasiparticle}. Such a dramatic contrast suggests that the \emph{mutual interplay} between the spin and charge degrees of freedom, instead of the pure ``RVB'' itself, may play a decisive role in the LE phase. Here the $\sigma\cdot$$t$-$J$ model is equivalent to a $t$-$J$ type model with a ferromagnetic (FM) in-plane spin superexchange \cite{jiang2019critical}. Moreover, although a long-range N\'{e}el ordered state would naturally occur in the two-dimensional (2D) case \cite{auerbach2012interacting}, such an order may be still realized  in the spin ladder case by reducing the in-plane (transverse) spin superexchange coupling close to the Ising limit \cite{narushima1995numerical,roy2017detecting,li2017groundstate}. Thus, by systematically studying a doped XXZ model on a two-leg ladder as a prototypical Mott insulator, one may gain a full understanding of the mutual spin-charge interaction with utilizing the precise numerical tool-DMRG \cite{white1992density,white1993density,schollwock2005density},  which can in turn shed important light on the realistic 2D physics \cite{anderson1987resonating} related to the high-$T_c$ cuprate \cite{bednorz1986possible,keimer2015quantum}.

\begin{figure}
  \centering
  \includegraphics[width=\linewidth]{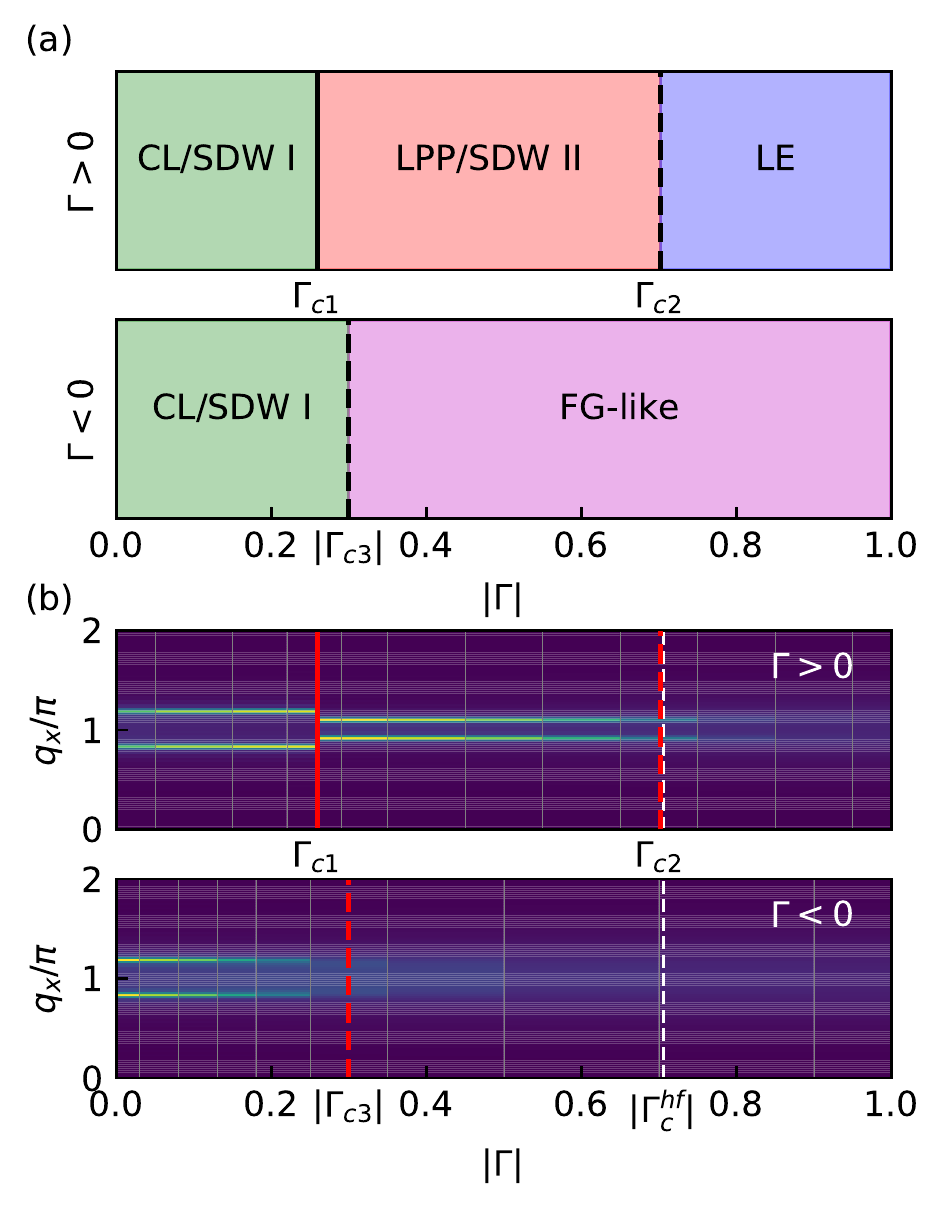}
  \caption{
    (a) The schematic phase diagram of finite-doped two-leg XXZ ladder. Upper panel at $\Gamma > 0$, two insulating phases separated by a solid line:  CL/SDW~\rom{1} denotes the charge localization with an SDW of wavelength $\lambda_{\text{SDW}} = 1/\delta$, while LPP/SDW~\rom{2} refers to the lower pseudogap phase with an SDW of wavelength $\lambda_{\text{SDW}} = 2/\delta$. The latter is further separated from a quasi-1D SC (LE) phase at  $\Gamma > \Gamma_{c2}$ by a dashed vertical line; Lower panel at $\Gamma < 0$: CL/SDW~\rom{1} persists to $\Gamma_{c3}$ (dashed vertical line) to reduce to a Luttinger liquid phase in the Fermi gas limit. 
   (b) The $z$-component spin structure factor $S^{zz}(q_{x}, q_{y})$ with $q_{y} = \pi$ at $\delta=1/12$ and $N_{x} = 48$, which shows sharp peaks at $\pi\pm 2\pi\delta$ and $\pi\pm \pi\delta$, respectively, indicating distinct SDW orders in SDW~\rom{1}  and SDW~\rom{2} separated by the red solid line at $\Gamma_{1c}$. The SDW orders disappears in the LE and FG-like phases, marked by red dashed lines at $\Gamma_{2c}>0$ and $\Gamma_{3c}<0$, respectively. Note that the white dashed lines at $|\Gamma^{hf}_{c}|$ highlight the phase boundary of the XXZ model at half-filling \cite{li2017groundstate}.
  }
  \label{fig:phas_diag}
\end{figure}

In this paper, we present a complex phase diagram obtained by DMRG in the lightly doped two-leg XXZ ladder via tuning the superexchange anisotropy at a given doping concentration. Near the Ising limit, a coexisting phase of charge localization (CL) and long-range spin density wave (SDW) (i.e., CL/SDW~\rom{1}) is identified. By reducing the anisotropy, the charge pairing and a distinct SDW with doubled wavelength emerges in a \emph{new} phase (i.e., LPP/SDW~\rom{2}). Here LPP refers to a so-called lower pseudogap phase where the pairing is present but the system is short of SC phase coherence \cite{weng2006lower}. Such a phase still remains insulating, concomitant with a classical SDW order in which the transverse (Goldstone-like) mode is gapped. Only when the long-range SDW order truly disappears at a weaker anisotropy, does the SC correlation finally become quasi-long-ranged as characterized by the Luther-Emery liquid which has been already well-established in the isotropic (t-J) limit \cite{jiang2018superconductivity,jiang2019critical}.  On the contrary, in the in-plane FM XXZ regime, no pairing between the holes is ever found, where a FG-like phase emerges after the CL/SDW~\rom{1} vanishes. A schematic phase diagram for such a lightly doped XXZ model is summarized in Fig.~\ref{fig:phas_diag}(a).

\emph{\label{sec:mm}Model and Method.---}The doped XXZ model on a two-leg square ladder is given by
\begin{equation}
  \label{eq:txxz_ham}
  H = H_{t} + H_{\mathrm{XXZ}}~.
\end{equation}
Here, $H_{t}$ describes the hole hopping process
\begin{equation}
  \label{eq:txxz_ham_t}
  H_{t} = -t\sum_{\langle ij \rangle \sigma}\hat{c}^{\dagger}_{i\sigma}\hat{c}_{j\sigma} +
  h.c.~,
\end{equation}
where $\langle ij \rangle$ denotes the nearest-neighbor sites, with the no-double-occupancy constraint always enforced such that only a single electron can occupy each site: $\hat{n}_{i} \leq 1$. The spin superexchange term is given by
\begin{equation}
  \label{eq:txxz_ham_xxz}
  H_{\mathrm{XXZ}} = J\sum_{\langle ij \rangle} \left[\hat{S}^{z}_{i}\hat{S}^{z}_{j} + \Gamma/2 (\hat{S}^{+}_{i}\hat{S}^{-}_{j} + \hat{S}^{-}_{i}\hat{S}^{+}_{j})\right]~,
\end{equation}
where $J$ is the strength of the $z$-component coupling, while $\Gamma$ denotes the anisotropy in the strength of the in-plane or transverse components. We shall fix $t/J =3$ and tune $\Gamma$ to study the evolution of the ground state at a doping concentration $\delta = N^{h}/N $ (typically at $\delta=1/12$ while other $\delta$'s have been also examined). Here $N^{h}$ is the number of holes and $N$ is the number of the lattice sites. Several limiting cases of the Hamiltonian (\ref{eq:txxz_ham}) have been investigated previously. At the isotropic $\Gamma = 1$ limit, the ground state is an LE liquid with quasi-1D SC \cite{troyer1996properties,hayward1996luttinger,muller1998phase,jiang2019critical} and it dramatically reduces to a Luttinger liquid behavior close to the free Fermi gas limit at $\Gamma = -1$ \cite{jiang2019critical}. At the Ising limit of $\Gamma = 0$, the doped holes repel each other with a vanished inverse compressibility which indicates the charge localization \cite{weisse2006t}.

We employ DMRG to determine the ground state of the Hamiltonian (\ref{eq:txxz_ham}) with system size $N=2\times N_x$ up to $N_x=192$.  We shall keep up to 12000 number of states in each DMRG block to reduce the truncation error smaller than $2\times 10^{-9}$ and perform typically about 30-200 sweeps to achieve a reliable convergence.  The high-performance matrix product state algorithm library GraceQ/MPS2 \cite{sun2019gqmps2} is used to perform the simulations. Some detailed scaling analyses are also presented in 
%Supplemental Material.
Appendix.

\begin{figure}
  \centering
  \includegraphics[width=\linewidth]{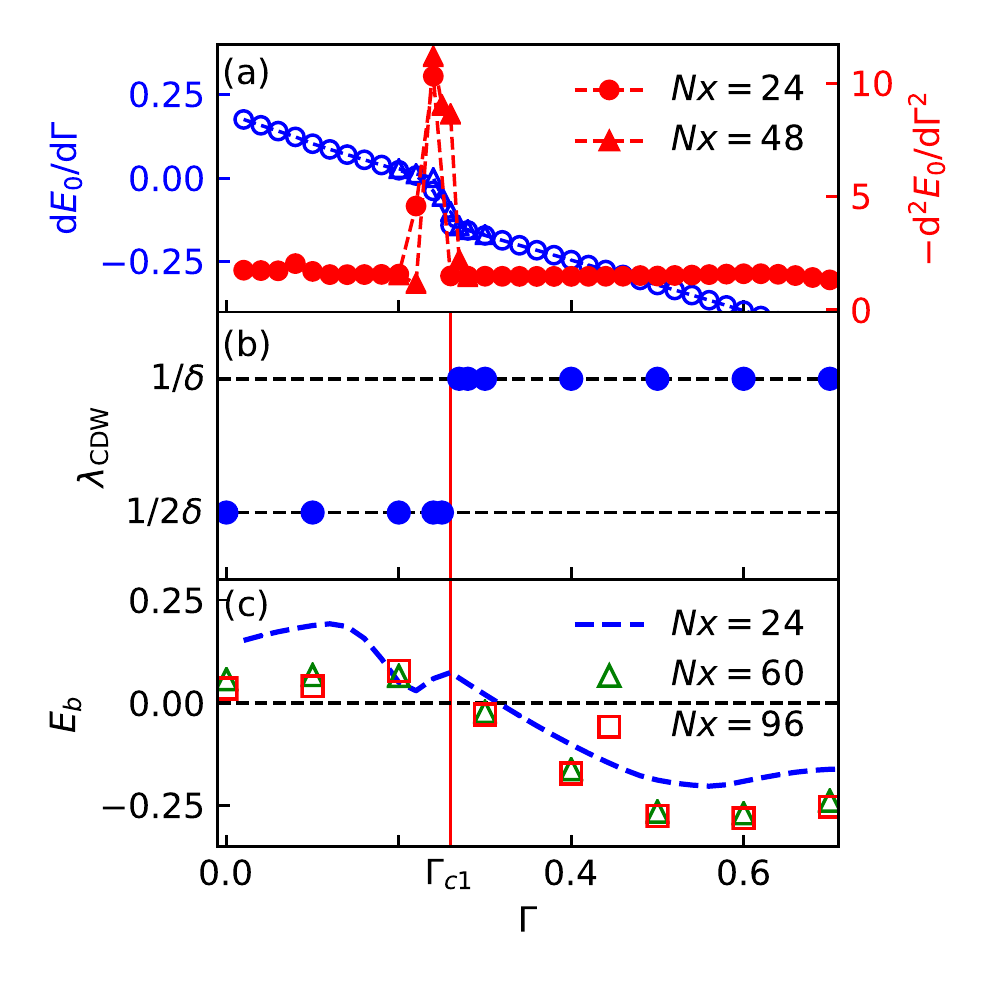}
  \caption{
    Two distinct SDW phases separated by a quantum critical point $\Gamma_{1c}$. (a) The first- and second-order derivatives of the ground state energy $E_0$ vs. $\Gamma$; (b) The CDW wavelength $\lambda_{\mathrm{CDW}}$, which exhibits a jump at $\Gamma_{1c}$ similar to $\lambda_{\mathrm{SDW}}$; (c) Binding energy $E_{b}$ is vanishingly small in CL/SDW~\rom{1} phase but scales to negative in LPP/SDW~\rom{2} phase. The red solid line highlights the phase boundary at $\Gamma_{1c}$ with $\delta=1/12$.
  }
  \label{fig:ploc2pp_e0div_cdwwl_eb}
\end{figure}
\begin{figure*}
  \centering
  \includegraphics[width=\linewidth]{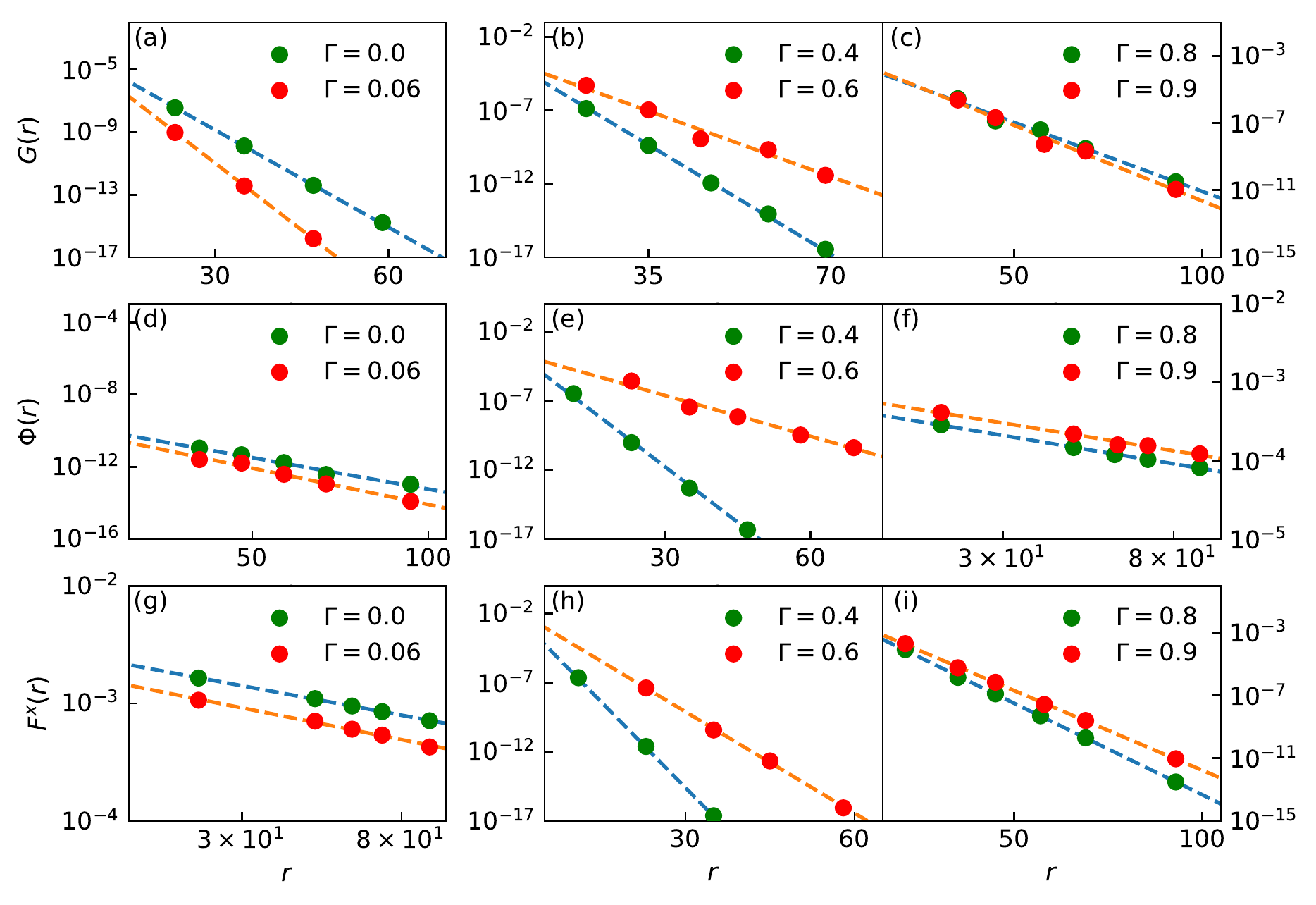}
  \caption{
    Characteristic correlation functions in three distinct phases at $\Gamma > 0$. Top panel: the single-particle Green's function $G_{\sigma}(r)$ in (a) CL/SDW~\rom{1}, (b) LPP/SDW~\rom{2}, and (c) LE. All in semi-logarithmic scale or $G_{\sigma}(r) \sim e^{-r/\xi_{G}}$;  Middle panel:  the pair-pair correlation function $\Phi(r)$ in (d) CL/SDW~\rom{1} and (e) LPP/SDW~\rom{2}, both in semi-logarithmic scale or $\Phi(r) \sim e^{-r/\xi_{sc}}$, and (f) LE in double-logarithmic scale indicating $\Phi(r) \sim r^{-K_{sc}}$; Bottom panel: the transverse spin-spin correlation $F^{x}(r)$ in (g) CL/SDW~\rom{1} in double-logarithmic scale, (h) LPP/SDW~\rom{2} in semi-logarithmic scale, and (i) LE phase in semi-logarithmic scale. For the detailed finite-size scaling analysis, see 
    %Supplemental Material.
Appendix.
  }
  \label{fig:collec_ntrvl}
\end{figure*}

\emph{\label{sec:corr}Correlation functions.---}Using the optimized DMRG ground state, the following physical quantities are investigated. The rung-averaged charge density along the ladder ($x$) direction is defined by $n(x)$, whose charge modulation behavior will be fitted by
\begin{equation}
  \label{eq:n_x_fit}
  n(x) = A_{\mathrm {CDW}}(N_{x})\cos(Q_{{\mathrm CDW}} \cdot x + \theta) + n_{0}~,
\end{equation}
where $Q_{\mathrm {CDW}}$ denotes the CDW wavevector and $A_{\mathrm {CDW}}$ the amplitude, which is scaled by $A_{\mathrm {CDW}}\sim N_{x}^{-K_{c}/2}$ with an exponent $K_{c}$. The rung-averaged equal-time single-particle Green's function is defined as
\begin{equation}
  \label{eq:cdagc}
  G_{\sigma}(r) = \frac{1}{2} \sum_{y = 0}^{1} | \langle
  \hat{c}^{\dagger}_{x_{0},y,\sigma}\hat{c}_{x_{0}+r,y,\sigma} \rangle|~,
\end{equation}
where $x_{0}$ labelled the reference rung. An exponentially decayed $G_{\sigma}(r) \sim e^{-r/\xi_{G}}$ is characterized by the correlation length $\xi_{G}$, while a power-law-decayed $G_{\sigma} \sim r^{-K_{G}}$ is fitted by a Luttinger exponent $K_{G}$. The rung-averaged pair-pair correlation is given by
\begin{equation}
  \label{eq:sc_corr}
  \Phi_{\alpha\beta}(r) = \frac{1}{2} \sum_{y = 0}^{1}|\langle
  \hat{\Delta}^{\dagger}_{\alpha}(x_{0}, y)\hat{\Delta}_{\beta}(x_{0}+r, y)\rangle|~,
\end{equation}
where $\hat{\Delta}^{\dagger}_{\alpha}(x, y) = \frac{1}{\sqrt{2}}\left[\hat{c}^{\dagger}_{(x,y),\uparrow}\hat{c}^{\dagger}_{(x,y)+\alpha,\downarrow} - \hat{c}^{\dagger}_{(x,y),\downarrow} \hat{c}^{\dagger}_{(x,y)+\alpha,\uparrow} \right]$ is the spin-singlet pair-field creation operator and $\alpha,\beta = \hat{x}, \hat{y}$ defines the pairing bond orientation.  The quasi-1D SC or non-SC state is characterized \cite{jiang2018superconductivity,jiang2019critical}  by $\Phi(r) \sim r^{-K_{sc}}$ or $\Phi(r) \sim e^{-r/\xi_{sc}}$, respectively.

Besides the above three quantities involving the charge degrees of freedom, the underlying spin degrees of freedom are described by the following quantities. The $z$-component spin structure factor is given by 
\begin{equation}
  \label{eq:szk}
  S^{zz}(\mathbf{q}) = \frac{1}{\sqrt{N}}
  \sum_{ij}e^{i\mathbf{q}\cdot(\mathbf{r}_{i}-\mathbf{r}_{j})}\langle
  \hat{S}^{z}_{i} \hat{S}^{z}_{j}\rangle~.
\end{equation}
The $z(x)$-component spin-spin correlations are defined by
\begin{equation}
  \label{eq:ss_corr}
  F^{z(x)}(r) = \frac{1}{2} \sum_{y = 0}^{1} |\langle (\hat{S}^{z(x)}_{x_{0},y} -
  \langle \hat{S}^{z(x)}_{x_{0},y} \rangle)(\hat{S}^{z(x)}_{x_{0}+r,y} -
  \langle \hat{S}^{z(x)}_{x_{0}+r,y} \rangle) \rangle|~.
\end{equation}
Their asymptotic behavior will be either captured by the correlation length scale $\xi_{F^{z(x)}}$ or a Luttinger exponent
$K_{F^{z(x)}}$ depending on the short-range or quasi-long-range order, respectively.

\emph{\label{sec:ploc2lpp}Two distinct insulating phases with SDW orders.---}Let us first focus on the phase diagram of Fig.~\ref{fig:phas_diag}(a) at $\Gamma>0$ (top row) and examine the two sides of the critical point at $\Gamma_{c1}$. As shown in the top row of Fig.~\ref{fig:phas_diag}(b), the $S^z$ structure factor clearly indicates that $\Gamma_{c1}$ separates two SDW ordered states with the wavevector $q_x$ peaked at $\pi\pm 2\pi\delta$ (SDW~\rom{1}) and $\pi\pm \pi\delta$ (SDW~\rom{2}) with the ``incommensurate'' wavelength $\lambda_{\mathrm {SDW}}=1/\delta$ and $2/\delta$, respectively. 

The first- and second-order derivatives of the ground state energy in Fig.~\ref{fig:ploc2pp_e0div_cdwwl_eb}(a) show a singularity at $\Gamma_{c1}$, suggesting a ``second-order-like'' phase transition. The wavelength of the CDW shown in Fig.~\ref{fig:ploc2pp_e0div_cdwwl_eb}(b) further demonstrates a jump at $\Gamma_{c1}$, which is consistent with that of the SDWs by $\lambda_{\mathrm {CDW}} =\lambda_{\mathrm {SDW}}/2$ (cf. Fig.~\ref{fig:phas_diag}(b) and
%Supplemental Material
Appendix
). The doubled CDW wavelength at $\Gamma>\Gamma_{c1}$ seems to indicate the pairing of the doped holes as opposed to the unpaired case at $\Gamma<\Gamma_{c1}$. 
Here we further examine the binding energy between the doped holes as defined by 
\begin{equation}
  \label{eq:eb}
  E_{b} = (E_{0}(N^{h} + 2) - E_{0}(N^{h})) - 2(E_{0}(N^{h} + 1) -
  E_{0}(N^{h}))~,
\end{equation}
where $E_{0}(x)$ is the ground state energy with $x$ holes.  As shown in Fig.~\ref{fig:ploc2pp_e0div_cdwwl_eb}(c), $E_{b}$ is always positive and extrapolated to vanishing at $\Gamma<\Gamma_{c1}$, but is clearly saturated  to a negative value at $\Gamma>\Gamma_{c1}$. 

However, the finite binding energy of holes does not necessarily imply a (quasi) long-range pair-pair correlation, which is absent on both sides of $\Gamma_{c1}$ as presented by the exponential decays in Figs.~\ref{fig:collec_ntrvl}(d) and (e). The corresponding single-particle Green's function is also exponentially decaying as shown in Figs.~\ref{fig:collec_ntrvl}(a) and (b). As far as the charge is concerned, two distinct phases separated by $\Gamma_{c1}$ seem all insulating, which is distinguished by the presence (absence) of local charge pairing. In Fig.~\ref{fig:phas_diag}(a), these two SDW phases are thus further denoted by CL and LPP.

One may further compare the transverse spin-spin correlation on either side of $\Gamma_{c1}$ in Figs.~\ref{fig:collec_ntrvl}(g) and (h), respectively. The quantum nature of the magnetic order in SDW~\rom{1} is supported by the power-law behavior in Fig.~\ref{fig:collec_ntrvl}(g), namely, the presence of the Goldstone-like mode in the transverse components. Note that the disorder-free ``auto-localization'' in the Ising ($\Gamma=0$) limit has been already pointed out in early analytical work \cite{bulaevski1968new}, which should continuously persist in the regime of SDW~\rom{1}. But in the SDW~\rom{2}, the transverse spin-spin correlation becomes exponentially decaying in Fig.~\ref{fig:collec_ntrvl}(h) as if the Goldstone-like mode be gapped via some novel ``Higgs mechanism'' concomitant with the pairing.  It also indicates that LPP/SDW~\rom{2} resembles a \emph{classical} magnetically ordered phase without the rigidity from the quantum protection of the gapless transverse mode.  

\emph{Phase coherence achieved in the ``spin liquid'' regime: Superconducting at $\Gamma>\Gamma_{c2}$ vs. normal Fermi liquid at $\Gamma<\Gamma_{c3}<0$.---}As illustrated in Fig.~\ref{fig:phas_diag}(a), the quasi long-range pair-pairing will be established at $\Gamma>\Gamma_{c2}$ as shown in Fig.~\ref{fig:collec_ntrvl}(f). Simultaneously the SDW order disappears [cf. Figs.~\ref{fig:phas_diag}(b)~and~\ref{fig:collec_ntrvl}(i)]. And the corresponding $G(r)$ decays exponentially [cf. Fig.~\ref{fig:collec_ntrvl}(c)]. This phase can be continuously connected to the well-studied LE phase in the isotropic limit at $\Gamma=1$ \cite{jiang2019critical}. In other words, the quasi-1D SC coherence can be only realized with the disappearance of the magnetic order even though the local pairing is already present in the SDW~\rom{2} phase.  Additionally, the establishment of the SC coherence at $\Gamma>\Gamma_{c2}$ can also be characterized by the qualitative change, from the exponential to the power-law decay, in the response to an inserted magnetic flux under periodic boundary condition \cite{seidel2005luther,zhu2014nature} which is presented in the
%Supplemental Material
Appendix
as a cross check.

\begin{figure}
  \centering
  \includegraphics[width=\linewidth]{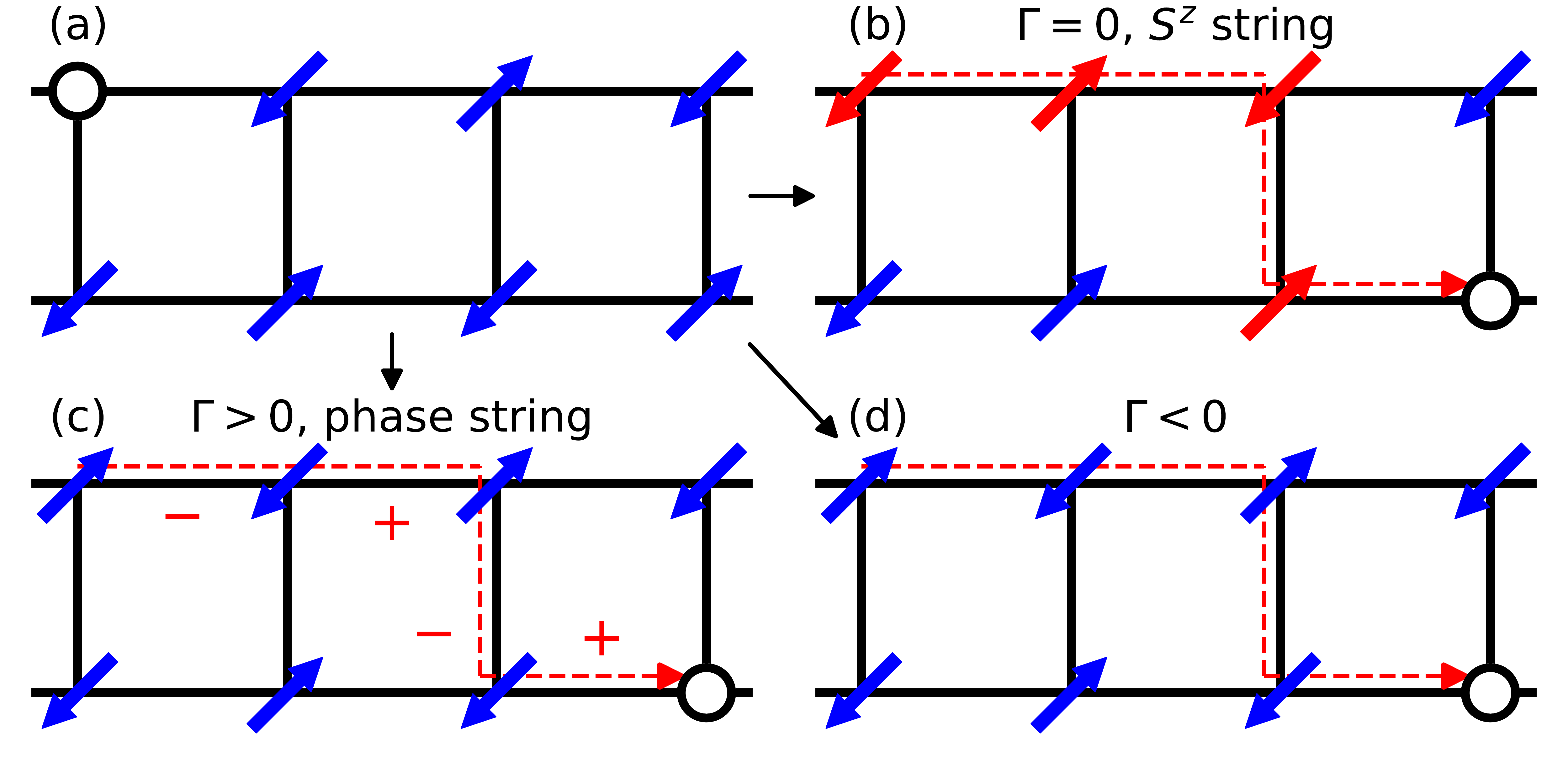}
  \caption{The underlying mechanism for the complex phase diagram: A doped hole in a N\'{e}el ordered state (a) will create an $S^{z}$-string in the Ising limit (b) to result in its self-localization, which may be ``erased'' via spin flips at sufficiently large $|\Gamma|$ [(c) and (d)]. However, in contrast to a \emph{complete} healing at $\Gamma<0$ in the FG-like phase, a transverse $S^{\pm}$-string defect known as the phase string will always be left behind at $\Gamma>0$, which can only be eliminated via the pairing of two holes, leading to the novel pairing mechanism in LPP/SDW~\rom{2} and LE phases.}
  \label{fig:sysskch}
\end{figure}

By comparison, at $\Gamma<0$, the SDW~\rom{1} order terminates at $\Gamma_{c3}<0$ as shown in Figs.~\ref{fig:phas_diag}(a) and (b). However, instead of pairing, the doped holes will form a \emph{coherent} Fermi-gas-like ground state  at $\Gamma<\Gamma_{c3}<0$, with the Luttinger exponent $K_{G} \sim  1$, as shown in
%Supplemental Material
Appendix
. It is emphasized that here the spin background also becomes a short-range ``spin liquid'' state as in the LE case at $\Gamma>\Gamma_{c2}$ since 
$H_{\mathrm{XXZ}}$ at $\Gamma<0$ can be exactly mapped to $H_{\mathrm{XXZ}}$ at $\Gamma>0$ by a unitary transformation \cite{jiang2019critical}. However, the hopping term $H_t$ is mapped to the hopping term of the so-called $\sigma\cdot$$t$-$J$ defined by
 $H_{\sigma\cdot t} = -t\sum_{\langle ij \rangle,\sigma}\sigma
 \hat{c}^{\dagger}_{i\sigma}\hat{c}_{j\sigma}+\mathrm {h.c.}$  \cite{jiang2019critical}.
Consequently the main difference between $\Gamma < 0$ and $\Gamma >0$ is precisely that between the $t$-$J$ and  $\sigma\cdot$$t$-$J$ models \cite{zhu2014nature,sun2019localization}, which can be solely attributed to the presence/absence of the phase string sign structure \cite{sheng1996phase,weng1997phase,wu2008sign} as to be further elaborated below.

 \emph{Discussion.---}Four distinct phases have been identified in the present lightly doped XXZ model. In the Ising limit $|\Gamma| \rightarrow 0$, the spins of the system form an SDW~\rom{1} order, where the doped holes are always \emph{self-localized} by the so-called $S^z$-string \cite{bulaevski1968new} (red-colored spins) as schematically illustrated in Fig.~\ref{fig:sysskch}(b), which are displaced by the hopping of holes to result in string-like mismatches of the FM pattern along the hole path. It cannot be ``repaired'' via transverse spin flips until $|\Gamma|$ is sufficiently large beyond the regime of $\Gamma_{3c}<\Gamma<\Gamma_{1c}$. At $\Gamma>\Gamma_{1c}$, the $S^z$-string does get erased via spin flips, but in general a sequence of signs (phase string)  \cite{sheng1996phase,weng1997phase,wu2008sign} will be acquired by the holes [Fig.~\ref{fig:sysskch}(c)], which is simply due to the fact that there is also a transverse $S^{\pm}$-string, similarly created by the hopping term $H_t$, which cannot be simultaneously ``repaired'' with the $S^{z}$-string through $H_{\mathrm{XXZ}}$. Any quantum spin zero-point motion can thus result in severe destructive interference effect due to Berry-phase-like phase strings associated with the holes. The only way to further eliminate such unescapable many-body quantum frustration (at least at low doping) is for holes to pair \cite{zhu2014nature,sun2019localization}  up in SDW~\rom{2} and LE phases at $\Gamma>\Gamma_{1c}$. In sharp contrast, at $\Gamma <0$, the model is equivalent to the  $\sigma\cdot$$t$-$J$ model, which does not generate any transverse spin string defect (phase string) and thus once $S^z$-string is erased at $\Gamma <\Gamma_{3c}<0$, the doped holes will simply behave like the free Landau's quasiparticles as illustrated in Fig.~\ref{fig:sysskch}(d), but there is no more incentive for holes to further pair up. 

\begin{acknowledgments}
Stimulating discussions with Shuai Chen, Hongchen Jiang, Donna Sheng and Jan Zaanen are acknowledged. RYS is grateful to Hongchen Jiang, Yifan Jiang, Cheng Peng for the GraceQ/MPS2 project and Sibo Zhao for computational resource management. This work is partially supported by Natural Science Foundation of China (Grant No. 11534007) and MOST of China (Grant No. 2017YFA0302902).
\end{acknowledgments}

\appendix*
\section{Supplemental Material}

\begin{figure*}
  \centering
  \includegraphics[width=\linewidth]{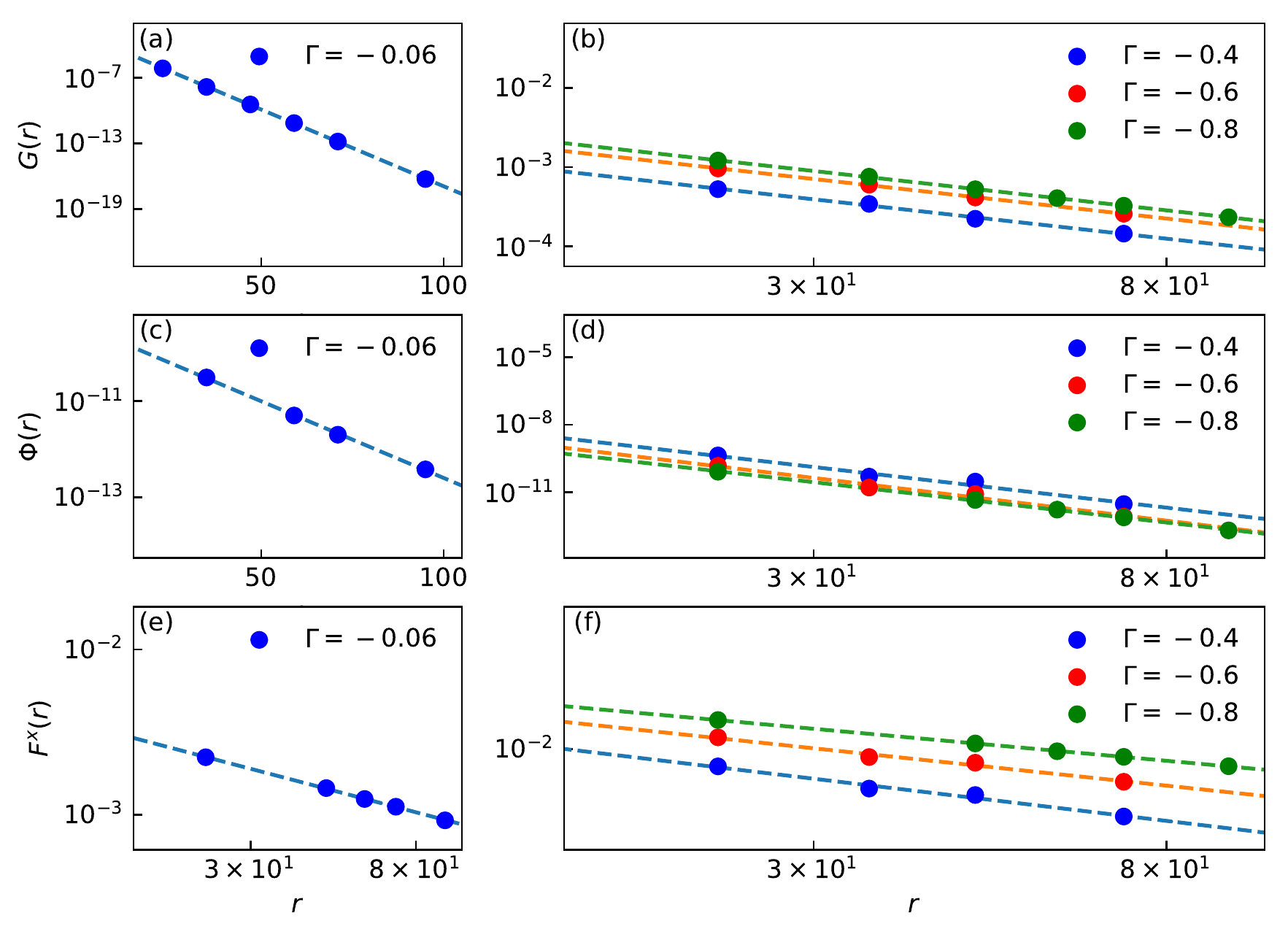}
  \caption{
  The typical correlation functions in two distinct phases at $\Gamma < 0$. Top panel: the single-particle Green's function $G_{\sigma}(r)$ in  (a) CL/SDW~\rom{1} phase in semi-logarithmic scale (i.e., $G_{\sigma}(r) \sim e^{-r/\xi_{G}}$),  and (b) FG-like phase in double-logarithmic scale with the leading Luttinger exponent $K_{G} \sim 1$. Middle panel: the pair-pair correlation function $\Phi(r)$ in (c) CL/SDW~\rom{1} phase in semi-logarithmic scale and (d) FG-like phase in double-logarithmic scale. Bottom panel: the transverse spin-spin correlation $F^{x}(r)$ in (e) CL/SDW~\rom{1} phase and (f) FG-like phase,  both of which in double-logarithmic scale.}
  \label{fig:collec_trvl}
\end{figure*}

\begin{figure*}
  \centering
  \includegraphics[width=0.8\linewidth]{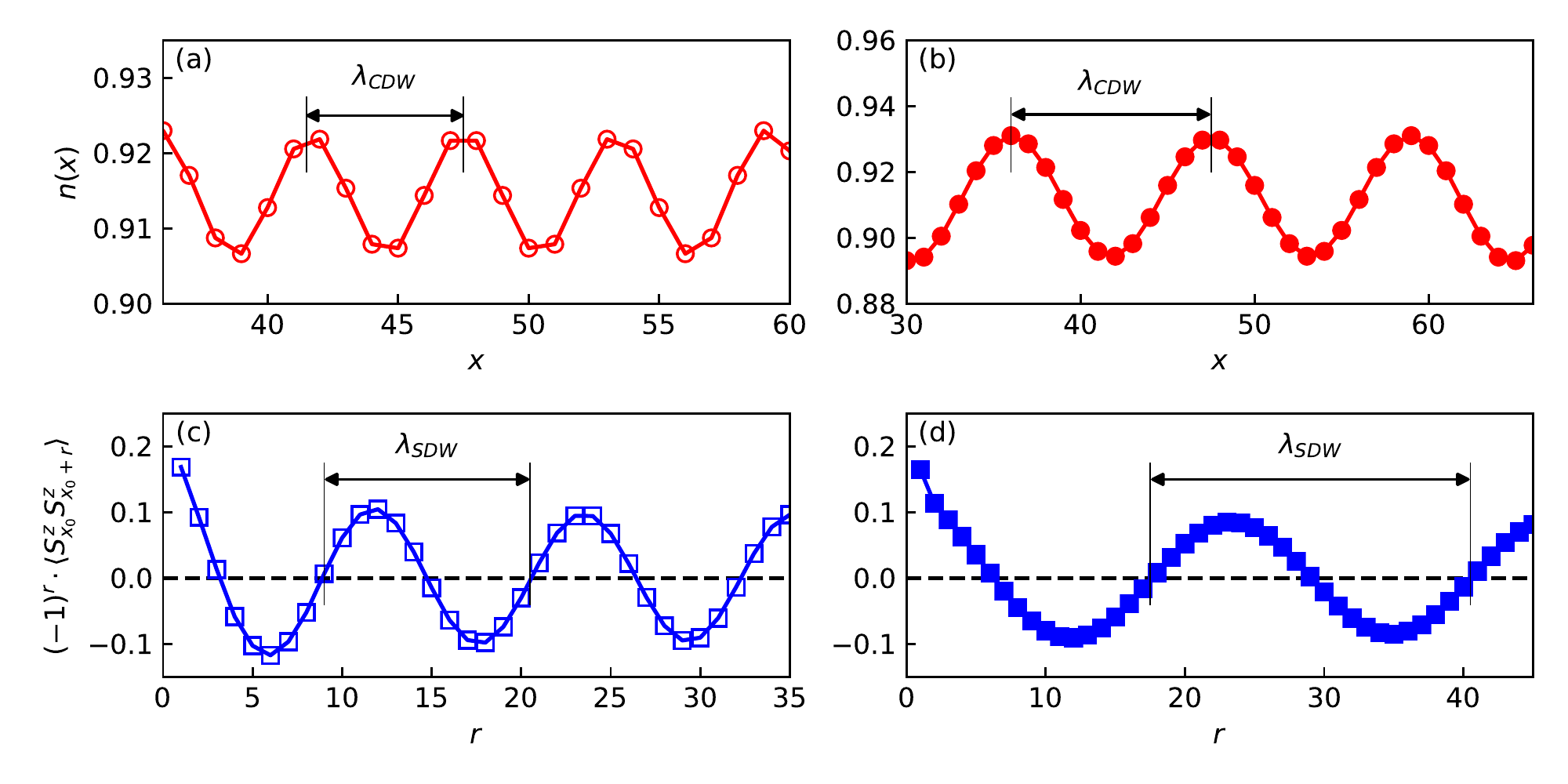}
  \caption{
  CDW and SDW profiles in the CL/SDW~\rom{1} phase and LPP/SDW~\rom{2} phase. (a) The CDW profile at $\Gamma = 0.06$ with $\lambda_{\mathrm{CDW}} = 1/(2\delta)$. (b) The CDW profile at $\Gamma = 0.6$ with $\lambda_{\mathrm{CDW}} = 1/\delta$. (c) The SDW profile at $\Gamma = 0.06$ with $\lambda_{\mathrm{SDW}} = 1/\delta$. (d) The SDW profile at $\Gamma = 0.6$ with $\lambda_{\mathrm{SDW}} = 2/\delta$. The system size is $N_{x} = 96$ and the reference point to measure SDW is at $N_{x} / 4$.
  }
  \label{fig:nx_sdw}
\end{figure*}

\begin{figure*}
  \centering
  \includegraphics[width=0.8\linewidth]{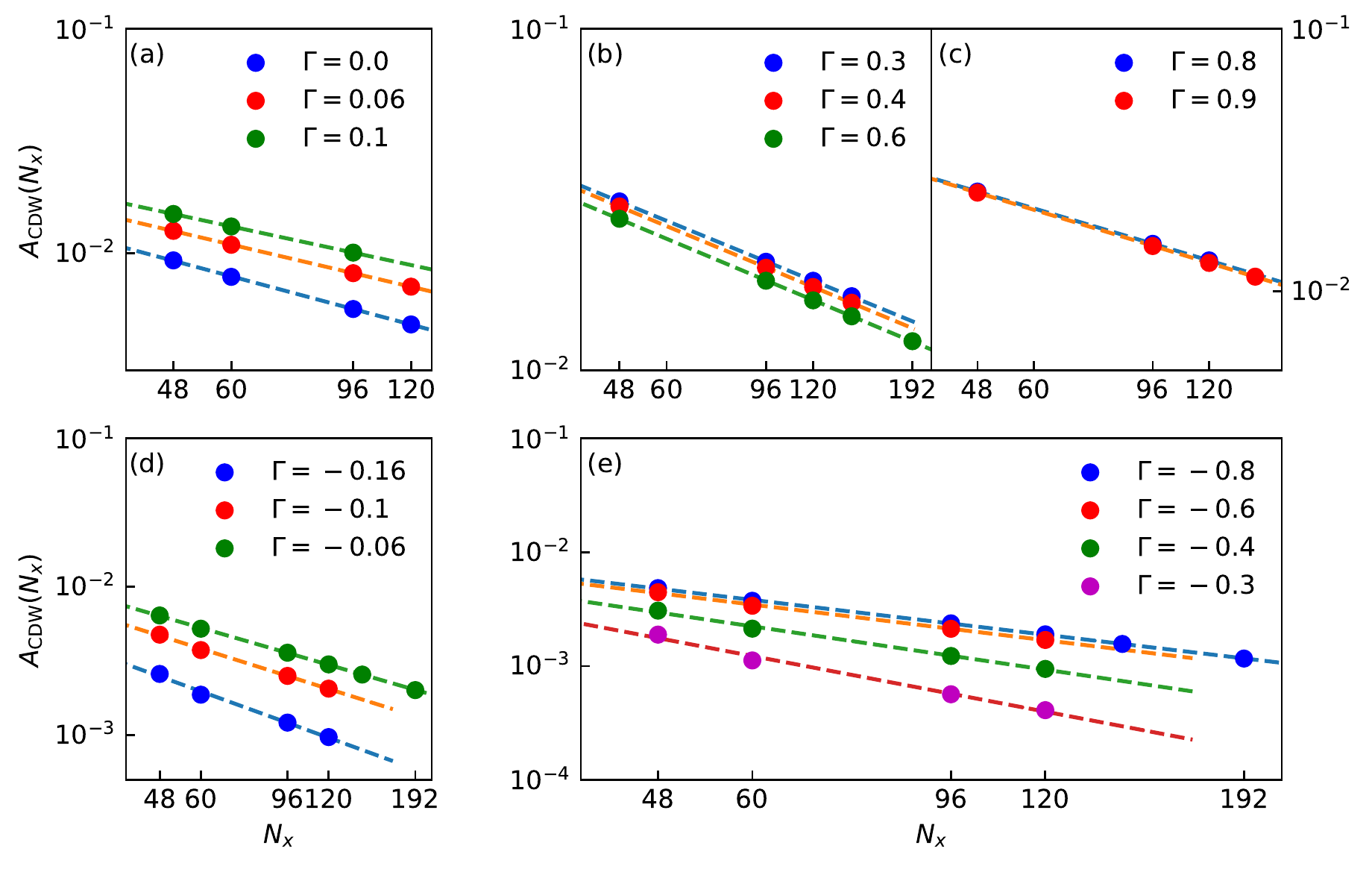}
  \caption{Finite-size scalings of CDW amplitude $A_{\mathrm{CDW}}$ with system length $N_x$ in double-logarithmic scale. Upper panels for $\Gamma >0$: (a) the CL/SDW~\rom{1} phase, (b) the LPP/SDW~\rom{2} phase, (c) the LE phase. Lower panels for $\Gamma < 0$: (d) the CL/SDW~\rom{1} phase, (e) the FG-like phase.}
  \label{fig:acdw}
\end{figure*}

\begin{figure*}
  \centering
  \includegraphics[width=0.8\linewidth]{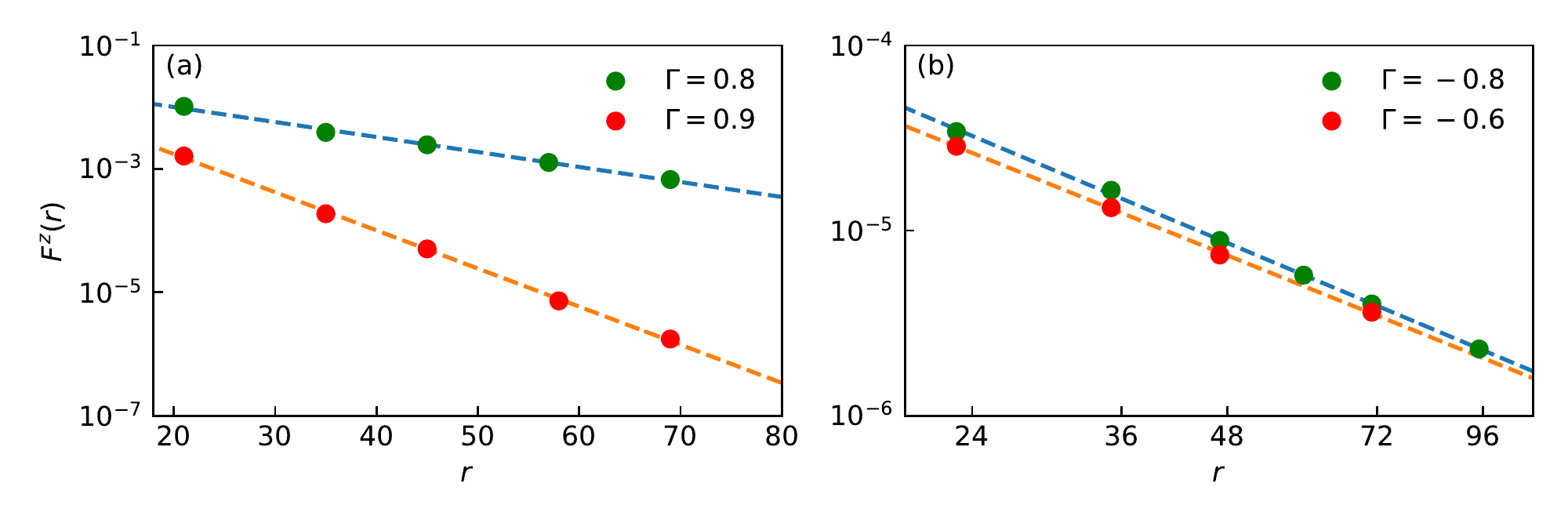}
  \caption{Finite-size scalings of $z$-component spin-spin correlation in the LE and FG-like phases. $F^{z}(r)$ in (a) LE phase in semi-logarithmic scale, or equivalently, $F^{z}(r) \sim e^{-r/\xi_{F^{z}}}$ , and (b) FG-like phase in double-logarithmic scale.}
  \label{fig:szsz}
\end{figure*}

\begin{figure}
  \centering
  \includegraphics[width=\linewidth]{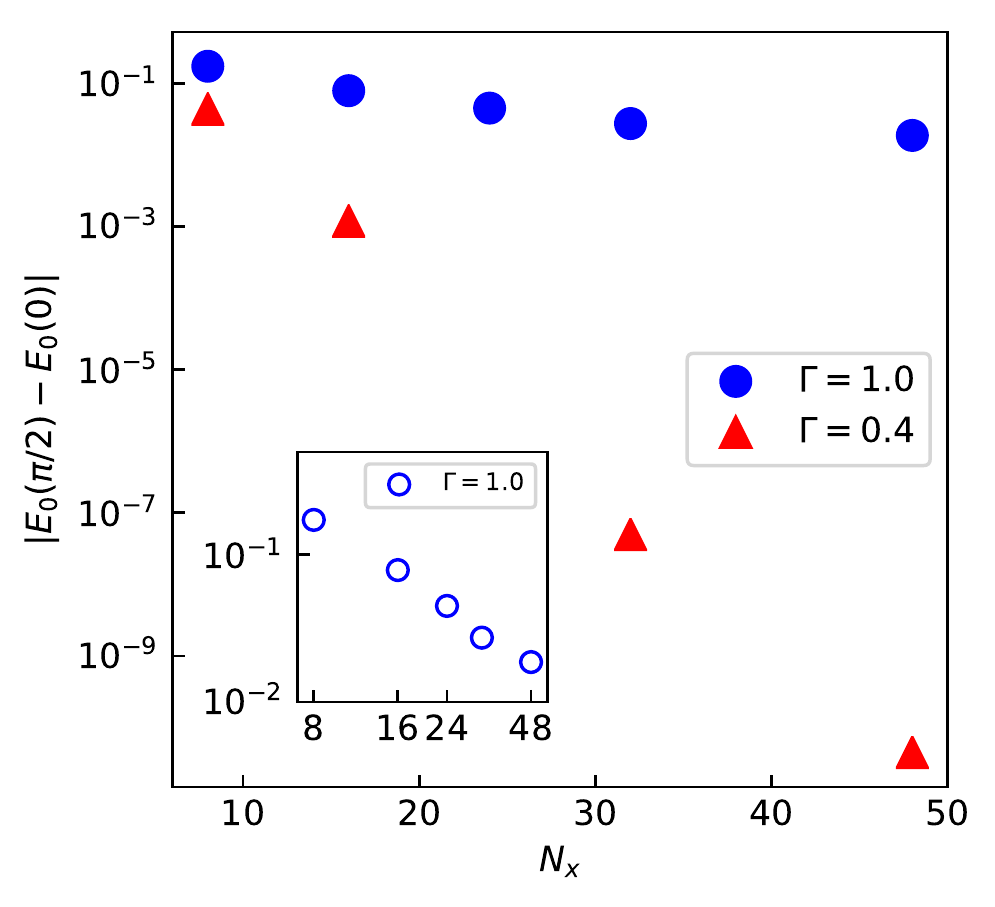}
  \caption{Finite-size scalings of the response of the ground state to a $\pi/2$ magnetic flux insertion. The blue circles represent the energy difference for a system at $\Gamma = 1.0$ in the LE phase and the red triangles represent the same quantity for a system at $\Gamma = 0.4$ in the LPP/SDW~\rom{2} phase in semi-logarithmic scale. The inset: rescale the $\Gamma = 1.0$ data in double-logarithmic scale. The doping concentration is fixed at $1/8$.}
  \label{fig:flux_response}
\end{figure}

This supplemental material contains more numerical details. We first introduce the finite-size scaling method used in this work, and then show the specific data points used to illustrate the spin structure factor in the main text. In the third section, we show additional numerical results.

\subsection{Finite-size scaling method}
In the recent works \cite{jiang2018superconductivity,jiang2019superconductivity,jiang2019critical}, a new finite-size scaling method is introduced to reduce the boundary effect, avoid the oscillation effect due to the intrinsic CDW, and increase the availability of sample size. The quintessence of this method is to measure the correlations between two CDW peaks to separate the CDW oscillation from the scaling of the envelop.  The sample length $N_{x}$ must be equivalent to the quadruple CDW period in the original version of this method, here we  slightly extend this method by relaxing the central $N_x/2$ restriction to any reasonable central distance between two CDW peaks. For example, for a system with $N_{x} = 60$ and $\lambda_{\mathrm{CDW}} = 12$, we can measure the correlations between two points across central three periods. The advantage of this extension is that more system sizes are accessible to measure correlation functions, while the disadvantage is that small measurement fluctuations may be induced by the boundary effect. All of the data points in this work are measured in the simulation case with minimal truncation error.

\subsection{Data details in spin structure factor illustration}
To illustrate the overall perspective of the $z$-component spin structure factor in the parameter space, we select discrete $\Gamma$ points to measure this quantity and broaden them to create a continuous picture in Fig. 1(b). The data points used to illustrate the upper panel of the Fig.~1(b) are: $\Gamma=$ 0.0, 0.1, 0.2, 0.24, 0.28, 0.3, 0.4, 0.5, 0.6, 0.7, 0.8, 0.9 and 1.0. The data points used to illustrate the lower panel of the Fig.~1(b) are: $\Gamma=$ 0.0, -0.06, -0.1, -0.16, -0.2, -0.3, -0.4, -0.6, -0.8, and -1.0. The boundaries of each broadening are marked using the gray solid lines.

\subsection{more numerical results}
In this section, we first show the characteristic correlation functions at $\Gamma < 0$. Then, we show more numerical results of CDW and SDW in the whole phase diagram. Then, we show the finite-size scaling results of the $z$-component spin-spin correlation function in LE and FG-like phases. Finally, we show the finite-size scaling results of the response of the ground state to a $\pi/2$ magnetic flux insertion.

\subsubsection{Correlation functions at $\Gamma < 0$}%
As a comparison with Fig.~3 in the main text, the same correlation functions on the $\Gamma < 0$ side are illustrated in Fig.~\ref{fig:collec_trvl}. 
For the regime near the $\Gamma = 0$ Ising limit, the system is still in the CL/SDW~\rom{1} phase. Figs.~\ref{fig:collec_trvl}(a),~\ref{fig:collec_trvl}(c)~and~\ref{fig:collec_trvl}(e) show the characteristic correlation functions, which exhibit the same behaviors as the ones at $\Gamma > 0$. At $\Gamma < \Gamma_{c3} < 0$, the leading Luttinger exponent $K_{G} \sim 1$ [cf. Fig.~\ref{fig:collec_trvl}(b)] indicates the holes form a \emph{coherent} Fermi-gas-like ground state. Both the pair-pair correlation [cf. Fig.~\ref{fig:collec_trvl}(d)] and transverse spin correlations [cf. Fig.~\ref{fig:collec_trvl}(f)]  decay in a power-law fashion but with a subleading exponent.

\subsubsection{CDW and SDW}

The CDW and SDW profiles in the CL/SDW~\rom{1} and LPP/SDW~\rom{2} phases are illustrated in Fig.~\ref{fig:nx_sdw}. They are characterized by the wavelengths, satisfying $\lambda_{\mathrm{SDW}} = 2\lambda_{\mathrm{CDW}}$. In the CL/SDW~\rom{1} phase, $\lambda_{\mathrm{CDW}}=1/(2\delta)$ is consistent with single holes carrying spin anti-domain walls along the $x$-direction, while $\lambda_{\mathrm{CDW}}=1/\delta$ in the LPP/SDW~\rom{2} phase is consistent with the pairing of two holes. The finite-size scaling results of the CDW amplitude $A_{\mathrm{CDW}}$ are illustrated in Fig. \ref{fig:acdw}. It is always in power-law decay with the sample size in the whole phase diagram.

\subsubsection{$F^{z}(r)$ in the LE and FG-like phases}%
\label{sub:fz_in_the_le_and_fl}

The $z$-component SDW order is absent in the LE and FG-like phases. The $z$-component spin-spin correlation function $F^{z}(r)$ in these two phases are illustrated in Fig.~\ref{fig:szsz}. The exponential decay of $F^{z}(r)$ [cf. Fig.~\ref{fig:szsz}(a)] displays the absence of the magnetic order,  suggesting the competition between superconductivity and SDW order. Although the $F^{z}(r)$ decays power-lawly in the FG-like phase as shown in Fig.~\ref{fig:szsz}(b), it has a subleading exponent, with the leading Luttinger exponent given by the single-particle Green's function of the doped holes. On the other hand, the spin background described by the XXZ term should be also in a short-ranged spin liquid similar to that in the LE state, as discussed in the main text. 

\subsubsection{Magnetic flux response in the LPP/SDW~\rom{2} and LE phases}
\label{sub:flux_response_in_the_lpp_and_le}
Except the pair-pair correlation function, we can also explore the response of a system with periodic boundary condition (PBC) to a magnetic flux which is inserted to the PBC ring to identify whether the pairs exhibit quasi-long-ranged coherence in the charge-paired LPP/SDW~\rom{2} and LE phases. If the charge $2e$ pairs can move coherently throughout the ring, then a $\pi/2$ magnetic flux can be picked up as a nontrivial Berry's phase to result in the energy difference showing a power-law behavior dut to the quasi-1D nature. On the contrary, for the incoherent pairs such a magnetic flux will not be felt in a sufficiently large-size ring such that the energy difference is expected to decay exponentially with the system size increasing \cite{seidel2005luther,zhu2014nature}. The magnetic flux responses in the LPP/SDW~\rom{2} and LE phases are illustrated in the Fig.~\ref{fig:flux_response}. The energy difference clearly decays exponentially in the LPP/SDW\rom{2} phase (cf. red triangles in the main panel). By contrast, it decays much slower in the LE phase (cf. blue circles in the main panel), and as a matter of fact, it exhibits a power-law behavior as shown in the inset.

\end{document}